\documentclass[useAMS,usenatbib]{mn2e}

\newcommand{\vect}[1]{\mathbf{#1}}   
\newcommand{\dif}[2]{\frac{{\rm d} #1}{{\rm d} #2}}
\usepackage{graphicx}
\usepackage{epstopdf}

\title[Ultra-High-Energy Cosmic Rays from the Radio Lobes of AGNs]{Ultra-High-Energy Cosmic Rays from the Radio Lobes of AGNs}
\author[F. Fraschetti and F. Melia]{F. Fraschetti$^{1,2}$\thanks{E-mail:
federico.fraschetti@cea.fr; melia@physics.arizona.edu} and F. Melia$^{3}$\\
$^{1}$LUTh, Observatoire de Paris, CNRS-UMR8102 and Universit\'e Paris VII,
5 Place Jules Janssen, F-92195 Meudon C\'edex, France.\\
$^2$Laboratoire AIM, CEA/DSM - CNRS - Universit\'e Paris Diderot,
Irfu/Service d'Astrophysique, F-91191 Gif sur Yvette C\'edex, France.\\
$^{3}$Department of Physics and
Steward Observatory, The University of Arizona, Tucson, AZ 85721, USA.}

\begin{document}

\date{Accepted 2008 September 19. Received 2008 September 19; in original form 2008 July 23}

\pagerange{\pageref{firstpage}--\pageref{lastpage}} \pubyear{2008}

\maketitle

\label{firstpage}

\begin{abstract}
\noindent In the past year, the HiRes and Auger collaborations have
reported the discovery of a high-energy cutoff in the ultra-high energy
cosmic-ray (UHECR) spectrum, and an apparent clustering of the highest
energy events towards nearby active galactic nuclei (AGNs). Consensus is
building that such $\sim 10^{19}$--$10^{20}$ eV particles are accelerated
within the radio-bright lobes of these sources, but it is not yet clear how
this actually happens. In this paper, we report (to our knowledge) the
first treatment of stochastic particle acceleration in such environments from first
principles, showing that energies $\sim 10^{20}$ eV are reached in $\sim
10^6$ years for protons.
However, our findings reopen the question regarding whether the high-energy
cutoff is due solely to propagation effects, or whether it does in fact
represent the maximum energy permitted by the acceleration process itself.
\end{abstract}


\section{Introduction}

Cosmic rays are energetic charged particles traveling throughout
the Galaxy and the intergalactic medium under the influence of various physical
processes, including deflection by magnetic fields, and collisions with
other particles along their trajectory. Their energy spectrum measured at
Earth is a steep (roughly power-law) distribution with logarithmic index
$\alpha\sim 2.6$--3, extending up to a few times $10^{20}$ eV. Other than
their spectrum, these particles are characterized by their angular distribution
in the sky, and by their mass composition.

A highly significant steepening in the UHECR spectrum was
reported by both the HiRes (Abbasi et al. 2008) and \citet{auger2008c}
collaborations. (Distinguished from their lower-energy counterparts,
UHECRs have energies in excess of 1 EeV $\equiv 10^{18}$ eV.) This result
may be a strong confirmation of the predicted Greisen-Zatsepin-Kuzmin (GZK)
cutoff due to photohadronic interactions between the UHECRs and low-energy
photons in the cosmic microwave background (CMB) radiation (Greisen
1966; Zatsepin \& Kuz'min 1966). Together with the measured low
fraction of high-energy photons in the CR distribution, this measurement
already rules out so-called top-down models, in which the UHECRs represent the
decay products of high-mass dark matter particles created in the early Universe
(Semikoz et al. 2007). The measured photon flux is also in conflict with
scenarios in which UHECRs are produced by collisions between cosmic strings 
or topological defects (Bluemer et al. 2008, Auger 2008b). On the other 
hand, such energetic particles may still be produced via astrophysical 
acceleration mechanisms (see Torres \& Anchordoqui 2004 and other references cited therein).

UHECRs are not detected directly, but through the showers they create in
Earth's atmosphere (see, e.g., Melia 2009). Depending on the energy and 
type of primary particle, the ensuing cascade has characteristics that 
allow the ground-based observatories to determine not only whether the 
incoming UHECR is a photon, but also its atomic number. It should be 
pointed out, however, that a determination of the primaries' 
composition strongly relies on an extrapolation of current
phenomenological hadronic interaction models, so it remains rather
uncertain. The \citet{auger2007} data confirmed the dominance
of protons in primary cosmic rays, though they also exhibit evidence for
a mixed composition extending to energies as high as $\sim$ 50--60 EeV,
with a higher atomic number $Z$, up to $Z \sim 26$ (Unger et al. 2007).

But the most telling indicator for the possible origin of these UHECRs is
the discovery by Auger (Auger 2008a) of their clustering towards nearby 
($\sim 75$ Mpc) AGNs along the supergalactic plane. The significance 
of this correlation has been further strenghtened by a more recent 
analysis which weights the AGN spatial distribution by their hard 
X-ray flux (George et al. 2008). This raises at least two questions:
(1) How are the UHECRs accelerated to such high energies? and (2) given
these nearby sources, is the sharp suppression of UHECRs in the last
decade of their observed energies really due solely to the GZK effect,
or does it signal a limitation to the acceleration efficiency?

Previous attempts at understanding how particles are accelerated to
EeV energies and beyond have generally been based on first-order Fermi
acceleration (see, e.g., Ostrowski 2008, and other references therein) 
within shocks created by blast waves like those in supernova remnants 
(Fatuzzo \& Melia 2003, Crocker et al. 2005). But this process is
subject to kinematic restrictions that inhibit the particles from 
actually reaching ultra-high energies (see, e.g., Nayakshin \& Melia 
1998, Gallant et al. 1999). Recent numerical simulations have shown 
that an increase in the Lorentz factor $\gamma$ of ultra-relativistic 
shock waves steepens the observed spectrum (Niemiec \& Ostrowski 2006) 
and reduces its high-energy cutoff.

For these reasons, it is not plausible for UHECRs to emerge from astrophysical
environments, such as supernova remnants, where first-order processes are dominant
so long as the shock velocity is super-Alfv\'enic, because they cannot even contain such
high-energy particles (Hillas 1984)---the gyration radius of particles with energy
$\sim 10^{20}$ eV for a typical galactic magnetic field
is much larger than the size ($<10$ pc) of these structures.

On the other hand, a second-order Fermi process (Fermi 1949)
can explain observational features not addressed by the first-order process,
as in the case of a supernova remnant itself (see Cowsik \& Sarkar 1984).
Moreover, stochastic particle acceleration through a gyroresonant
interaction with MHD turbulence (a second-order Fermi process; see Fermi 1949) can be
very efficient if the Alfv\'en velocity approaches $c$ (Dermer \& Humi 2001).
The stochastic acceleration of particles by turbulent plasma waves has already
received some attention in the literature (see Liu et al. 2004, 2006, and references
cited therein, and Wolfe \& Melia 2006). Indeed, the feasibility of second-order 
Fermi acceleration in radio galaxies has been demonstrated through the steady 
re-acceleration of electrons in certain hot spots (Almudena Prieto et al. 2002).

Our treatment from first
principles, however, avoids many of the previously encountered unknowns and
limitations. In this paper, we report (to our knowledge) the first treatment of
stochastic acceleration of charged particles in the lobes of radio-bright AGNs
by directly computing the trajectory of individual particles. An earlier version
of this treatment---for the propagation of charged particles assumed already 
accelerated at TeV energy through the turbulent magnetic field 
at the Galactic centre---may be found in Ballantyne et al. 2007, 
and Wommer, Melia \& Fatuzzo 2008; by contrast, in the present paper 
both the propagation and acceleration are taken into account. We show 
that random scatterings (a second-order Fermi process) between the charges 
and fluctuations in a turbulent magnetic field can accelerate these particles 
up to ultra-high energies, provided a broad range of fluctuations is present 
in the system.

\section{Description of the model}

In our treatment, we follow the three-dimensional motion of {\it individual}
particles within a time-varying turbulent magnetic field. By avoiding the use
of equations describing statistical averages of the particle distribution, we
mitigate our dependence on unknown factors, such as the diffusion coefficient.
We also avoid such limitations as the Parker approximation (Padmanhaban 2001)
in the transport equation. However, a remaining unknown is the partitioning
between turbulent and background fields. For simplicity, we take the minimalist
approach and assume that the magnetic energy is divided equally between the two
components.

Another unknown is the turbulent distribution. For many real astrophysical plasmas,
the magnetic turbulence seems to be in accordance with the Kolmogorov spectrum.
This is seen, e.g., in the solar wind (Leamon et al. 1998) and through interstellar
scintillation (Lee \& Jokipii 1976); a more recent numerical analysis of MHD
turbulence confirms the general validity of the Kolmogorov power spectrum (Cho et
al. 2003). In addition, renormalization group techniques applied to the analysis of MHD
turbulence also favour a Kolmogorov power spectrum (for more details, see
Smith et al. 1998, and Verma 2004).

We model the radio lobe of an AGN as a sphere of radius ${\mathcal R}$,
a second parameter in our simulations.
A population of relativistic particles of mass $m$, protons or heavy ions,
with an energy $E = \gamma mc^2$, where $\gamma$ is the Lorentz factor,
is released in an inner sphere of radius ${\mathcal R'} \sim \alpha {\mathcal R}$.
The value of $\alpha$ must be much smaller than 1, otherwise very few particles
reach an energy $E >10^{18}$eV. For a small value of $\alpha$, the gyration
radius becomes comparable to the size of the acceleration region at $E >10^{18}$eV,
and therefore changes in $\alpha$ do not significantly alter the result.
For the sake of specificity, we use a value $\alpha \sim 10^{-3}$ in this paper.
Once released, the particles propagate through the turbulent field until they
escape the accele\-ration region and enter intergalactic space.

\subsection{Time varying turbulent field}

We use the \citet{gj94} prescription for gene\-rating the turbulent magnetic
field. Their principal aim of propagating individual particles through a magnetostatic
field was to compute the Fokker-Planck coefficients for a direct
comparison with analytic theory. For our purpose, we modify that prescription
to include a time-dependent phase factor that allows for temporal variations.

The global magnetic field is written as a sum of a background term $\vect{B}_0$,
constant and uniform, and a turbulent field varying in space and time (i.e.,
as a superposition of Alfv\'en waves).

The equation of motion of a relativistic test particle with charge $e$ and mass
$m$ moving in an electromagnetic field $F^{\mu \nu}$ is the Lorentz equation
(Landau \& Lifchitz 1975, Melia 2001)
\begin{equation}
mc \frac{d u^\mu}{ds} = \frac{e}{c} F^{\mu \nu} u_\nu
\label{geo}
\end{equation}
(with $\mu = 0, 1, 2, 3$), where $c$ is the speed of light in vacuum, $u^\mu
=\left(\gamma, \gamma {\vect{v}}/{c}  \right)$
is the four-velocity of the particle, $\gamma = 1/ \sqrt{1-(v/c)^2}$
is the Lorentz factor, and $s/c$ is the proper time.
We calculate the trajectory of the particle in a ma\-gnetic field
$\vect{B}(t, \vect{r}) = (mc/e)\vect{\Omega}(t, \vect{r})$
as a solution of the space components ($\mu = 1,2,3$) of Equation (1)
\begin{equation}
\frac{d\vect{u}(t)}{dt} = \delta \vect{\cal E}(t, \vect{r}) + \frac{\vect{u}(t)
\times\vect{\Omega}(t,\vect{r})}{\gamma(t)}\;,
\label{lorentz}
\end{equation}
where $t$ is the time in the rest frame of the acceleration region.
The quantity $\vect{\Omega}(t, \vect{r})$ in Equation (\ref{lorentz}) is given by
\begin{equation}
\vect{\Omega}(t, \vect{r}) =
\vect{\Omega}_0+\delta \vect{\Omega}(t, \vect{r})\;,
\end{equation}
where $\vect{\Omega}_0 \equiv (e/mc)
\vect{B}_0$, in terms of the background magnetic field $\vect{B}_0$,
and $\delta \vect{\Omega}(t, \vect{r})$ is the time-dependent turbulent
magnetic field. We ignore any large-scale background electric fields---a
reasonable assumption given that currents would quench any such fields
within the radio lobes of AGNs. The time variation of the magnetic field,
however, induces an electric field $\delta \vect{\cal{E}}(t, \vect{r}) 
\equiv (e/mc) \vect{E}(t, \vect{r})$ according to Faraday's law.

The procedure of building the turbulence calls for the random
generation of a given number $N$ of transverse waves $\vect{k}_i$, $i=1,..,N$ at every point of
physical space where the particle is found, each with a random amplitude, phase
and orientation defined by angles $\theta (k_i)$ and $\phi(k_i)$. This form of the
fluctuation satisfies $\nabla \cdot \delta \vect{\Omega}(t, \vect{r})=0$. We write
\begin{equation}
\delta \vect{\Omega}(t, \vect{r}) = \sum_{i=1}^N\Omega(k_i) \hat{\xi}_{\pm} (k_i)
e^{\left[i(k_i x' - \omega_i t +
\beta(k_i))\right]}\;,
\label{omega}
\end{equation}
where the polarization vector is given by
\begin{equation}
\hat\xi_{\pm} (k_i)= \cos\alpha(k_i)\hat{\bf y}' \pm i\sin\alpha(k_i)\hat{\bf z}'\;.
\end{equation}
Given the form in Equation (4) for the turbulence, the electric field $\delta \vect{\cal{E}}(t, \vect{r})$ is given by
\begin{equation}
\delta \vect{\cal{E}}(t, \vect{r}) =
\sum_{i=1}^N\Omega(k_i) \frac{\omega(k_i)}{k_i c} \hat\xi_{\pm} ^E (k_i)
e^{\left[i(k_i x' - \omega_i t +
\beta(k_i))\right]}\;,
\label{electric}
\end{equation}
with
\begin{equation}
\hat\xi_{\pm} ^E (k_i)= \pm i\sin\alpha(k_i)\hat{\bf y}' - \cos\alpha(k_i)\hat{\bf z}'\;.
\end{equation}
The orthonormal primed coordinates ${\bf r'}  = (\hat{x}',\hat{y}',\hat{z}')$
are related to the lab-frame coordinates ${\bf r}  = (\hat{x},\hat{y},\hat{z})$
via the rotation matrix $R(\theta,\phi)$, in such a way that for every $k$ the
propagation vector is parallel to the $\hat{x}'$ axis. The matrix $R(\theta,\phi)$ is given by
\begin{equation}
{\bf r'} =
\left(
\begin{array}{ccc}
\cos\theta \; \cos\phi &\cos\theta \; \sin\phi  &\sin\theta \\
-\sin\phi  &\cos\phi  &0 \\
-\sin\theta \; \cos\phi  &-\sin\theta \; \sin\phi  &\cos\theta \\
\end{array}
\right) {\bf r}
\;.
\end{equation}
For each value of $k_i$, there are 5 random numbers: $0<\theta(k_i)<\pi$, $0<\phi(k_i)<2\pi$, $0<\alpha(k_i)<2\pi$,
$0<\beta(k_i)<2\pi$ and the sign plus or minus indicating the sense of polarization.

Further assumptions are necessary to specify the dispersion relation $\omega=\omega(k_i)$.
For every turbulent mode, we use the dispersion relation for transverse non-relativistic 
Alfv\'en waves (see Kaplan \& Tsytovich 1973 for an extended discussion): 
$\omega(k_i) = v_A k_i \cos\theta(k_i)$, for $i=1,..,N$,
where $v_A =  B_0/\sqrt{4\pi m_p n}$ is the non-relativistic
Alfv\'en velocity in a medium with background magnetic field $B_0$
and number density $n$, $m_p$ the proton mass, and $\theta(k_i)$ is
the angle between the wavevector $\vect{k}_i$ and $\vect{B}_0$. This is the condition
thought to be valid for the propagation of turbulent modes in a magnetized
astrophysical environment, such as the radio lobes of an AGN. The background
plasma is assumed to have a background proton number density $n \sim 10^{-4}$
cm$^{-3}$, a reasonable value for these environments (Almudena Prieto et al. 2002).

The amplitudes of the magnetic fluctuations are assumed to be
consistent with Kolmogorov turbulence, so
\begin{equation}
\Omega(k_i) = \Omega(k_{min}) \left(\frac{k_i}{k_{min}}\right)^{-\Gamma/2}\;,
\end{equation}
for $i=1,..,N$, where $k_{min}$ corresponds to the longest wavelength of the fluctuations and
the index $\Gamma$ of the power spectrum $\Omega^2(k)$ is $5/3$. Finally, the
quantity $\Omega(k_{min})$ is computed by requiring that the energy density of
the magnetic fluctuations equals that of the background magnetic field:
\begin{eqnarray}
\lefteqn{
S = \sum_{i=1}^N \frac{B^2(k_i)}{8\pi} = \frac{m^2 c^2}{8 \pi e^2} \Omega^2 (k_{min})
\sum_{i=1}^N \left(\frac{k_i}{k_{min}}\right)^{-\Gamma}= \frac{{B_0}^2}{8 \pi}.} \nonumber\\
\label{sum}
\end{eqnarray}
We choose N=2400 values of $k$ evenly spaced on a logarithmic scale;
i.e., a wavenumber shell with bounds
$k_i - k_{i+1}$ holds $k_{i+1} = k_i \times (k_{max}/k_{min})^{1/N}$ values.
Considering that the
turbulence wavenumber $k$ is related to the turbulent length scale $l$ by $k = 2\pi / l$,
we adopt a range of lengthscales from $l_{min} = 10^{-1}\,v_0/\Omega_0$ to $l_{max} =
10^{9}\,v_0/\Omega_0$, where $v_0$ is the initial velocity of the
particle and $\Omega_0$ is the initial gyrofrequency in the background magnetic field.
Thus the dynamic range covered by $k$ is $k_{max} / k_{min} = l_{max} / l_{min} = 10^{10}$,
and our description allows for $240$ transverse modes $k$ per decade.
The values of $k_{max}$ and $k_{min}$ fix the magnetic energy equipartition
through Equation (10). The value ${k_{min}} ^{-1}$ is proportional through a factor of order $1$
to the correlation length of the turbulence (see Ruffert and Melia 1994, and Rockefeller
et al. 2004, for examples of how this is generated in the interstellar medium); 
the value ${k_{max}} ^{-1}$ is the wavelength at which the interaction between 
the turbulence and most of the particles is the most efficient, so that energy 
is drained out of $\delta \Omega$.

However, since the gyroradius $r_g (E)$ evolves over
a large energy interval, the gyroresonant wavenumber $k_{res} (E)$ moves accordingly
in such a way that in the global wavenumber interval $(k_{min} - k_{max})$ there is
for every $E$ a certain $k_{res} (E)$ fulfilling the resonance condition $r_g (E) k_{res} (E) \sim 1$.
Such a $k$ range involves a large computational time,
especially if a statistically significant number of particles is to be considered.

Since our numerical simulation is not performed by specifying the magnetic turbulence
on a computational grid with given cell size $\Delta x$, the choice of $k_{max} = 2
\pi / l_{min}$ is not dictated by a fixed spatial resolution (see section \ref{num}
for more details). In addition, the result is not affected by spurious effects to the
discreteness or the periodicity. As a bypro\-duct, the divergenceless condition
$\nabla \cdot \delta \vect{\Omega}(t, \vect{r})=0$ is easily satisfied and does not
require an extension of the Godunov solver of the MHD equations for the purpose
of ``divergence cleaning'' (Ryu et al. 1998) or a reformulation of the MHD
equations including, e.g., divergence-damping terms (Dedner et al. 2002).

With this prescription, we construct the turbulent magnetic field at every point
of physical space where the particle is found, which we then propagate
without taking any time-average along the trajectory. The particles passing through
this region are released initially at a random position inside the acceleration
zone, which for simplicity is taken to be a sphere of radius ${\mathcal R}$, with
a fixed initial velocity $v_0$ pointed in a random direction. The initial value of the
Lorentz factor $\gamma_0 = 1/\sqrt{1 - (v_0/c)^2} \simeq 1.015$ is chosen to avoid having
to deal with ionization losses for the protons or heavy ions. The particles closest to
the edge of the acceleration zone have a higher probability of escaping than
those starting farther in, and therefore reach relatively lower energies. In the
usual (Fermi) way, this produces (in the highest energy portion of the spectrum)
an inverted power-law distribution.

\subsection{Energy losses}

In principle, energy losses due to synchrotron and inverse Compton processes
involving radio and Cosmic Microwave Background (CMB) photons, all of which increase
as $\gamma^2$, can significantly limit the maximum energy attainable by a
cosmic ray during the acceleration process, given that its Lorentz factor
$\gamma$ evolves from $\sim 1$ up to $10^{10}-10^{11}$. For an UHE particle
(either a proton or a heavy ion), both the radio and CMB photons will have
an energy $\gamma h \nu$ in the centre-of-momentum frame well below the rest
energy of the cosmic ray (i.e., $\gamma h \nu << m c^2$, where $m$
is the mass of the accele\-rating particle). For the purpose
of these estimates, we use a radio frequency $\nu_{radio} = 0.1$ GHz
($h \nu_{radio} \sim 4.2 \times 10^{-7}$ eV) and
a CMB frequency corresponding to the peak of the blackbody spectrum,
$\nu_{CMB} = 2.821 kT/h = 158$ GHz ($h \nu_{CMB} \sim 6.6 \times 10^{-4}$ eV),
where $k$ is the Boltzmann constant,
$h$ the Planck constant, and $T = 2.7$ K is the CMB temperature. Consequently,
the energy losses due to inverse Compton may be calculated in the Thomson limit.
Compare this with the situation for high energy electrons, for which the Thomson
condition would not be satisfied even at energies $\gamma m_e c^2 \sim 10^{16}$ eV,
requiring in that case the full Klein-Nishina treatment.

The propagation of high-energy particles is here mo\-deled in a region of tens of kpc size.
Therefore we neglect any effect of the relativistically-narrowed jet on the spatial distribution
of the radio background, assumed for simplicity to be isotropic.
Since the CMB intensity field is also isotropic, we take these
energy losses into account using the following angle-integrated power-loss rate:
\begin{equation}
-\frac{dE}{dt} = \frac{4}{3} \sigma_T(m) c \gamma^2
\left(\frac{B^2}{8\pi} + U_{R} + U_{CMB}\right)\;,
\label{loss}
\end{equation}
where $\sigma_T(m) = 6.6524 \times (m_e/m)^2\, 10^{-25}$ cm$^2$ is the Thomson
cross section for a generic particle of mass $m$, which can be a proton or heavy ion,
and $B^2/(8\pi) = (2 {B_0}^2)/(8\pi)$ is the total energy density of the
magnetic field. The photon energy density $U_{R}$ inside a typical radio
lobe is computed as $U_R = L/(4\pi c {\mathcal R}^2)$,
where we assume $L$ to be a standard luminosity density corresponding to
the Fanaroff-Riley class II of galaxies ($L = 5\times10^{25}$
W Hz$^{-1}$ sr$^{-1}$ at $178$ MHz), and ${\mathcal R}$
is the size of the spherical acceleration zone.
For the CMB, we use $U_{CMB} = a T^4
= 4.2 \times 10^{-13}$ erg cm$^{-3}$. In a region where magnetic turbulence
is absent or static, a given test particle propagates by ``bouncing" randomly
off the inhomogeneities in $\vect{B}$, but its energy remains constant. The
field we will model below, however,
is comprised of time-varying gyroresonant turbulent waves
(see Equation \ref{omega}), and collisions between the test particle and these
waves produces a net acceleration (in the lab frame).

\begin{figure}
\begin{center}
\includegraphics[width=9cm]{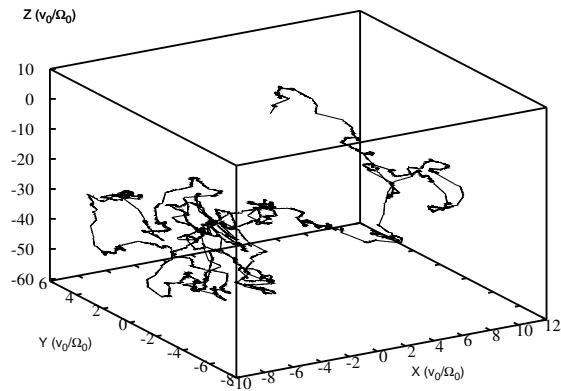}
\caption{Three-dimensional trajectory of single particle in the turbulent field
of \citet{gj94} reproduced with our code. The length scales are in units of the
gyration radius, $v_0 / {\Omega_0}$, which remains constant during the propagation.
The energy is verified to be constant, as expected, over a time interval
$\Delta t  = 1000 {\Omega_0}^{-1}$, within a relative error of $10^{-5}$.
The background magnetic field $B_0$ is parallel to the $z$ axis and, as found by
\citet{gj94}, the diffusion along ${\bf B}_0$ dominates with respect to that across
the field.}
\label{GJ}
\end{center}
\end{figure}

\begin{figure}
\begin{center}
\includegraphics[width=9cm]{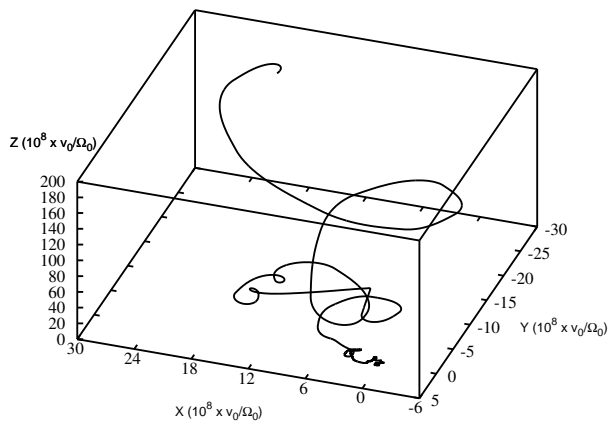}
\caption{Three-dimensional trajectory of a single particle
released at random within the acceleration zone, assumed to be a sphere of
radius 50 kpc. The scale on the axes is in units of $10^8 v_0/\Omega_0$,
where $v_0/\Omega_0$ is the initial gyroradius.
The particle is released with a fixed initial speed $v_0$, but pointed in
a randomly chosen direction.
The calculation stops when the particle leaves the radio lobe and
is injected into the intergalactic medium.}
\label{part}
\end{center}
\end{figure}

\section{Numerical code setup}\label{num}

In this section we describe the numerical code used to perform the simulations.
The Lorentz Equation (\ref{lorentz}) is integrated using a Runge-Kutta $4^{th}$
order method (Press et al. 1997) for the system of 6 first-order differential equations
\begin{eqnarray}
\dif{{\bf x_i}}{t} & = & \frac{c}{\gamma}{\bf u_i}\\
\frac{d\vect{u_i}}{dt} & = & \delta \vect{\cal{E}}_i + \frac{[\vect{u}
\times\vect{\Omega}]_i}{\gamma}\;,
\end{eqnarray}
for $i=1,2,3$. The components $\delta \vect{\cal{E}}_i$ and $\vect{\Omega}_i$ are
intended to be the real parts of the corresponding complex quantities.

The portable random number generator used to produce the turbulence is
Knuth's subtractive routine {\it ran3} (Press et al. 1997),
with a seed number $I=10^9$. This routine has a relatively
short execution time and is suitable to avoid the introduction
of unwanted correlations into the numerical computation.

There are two approaches to numerically implementing a turbulent magnetic field
generated by this method.
The first approach (used in Giacalone \& Jokipii 1994) is to calculate the magnetic
field at every time step for each particle position.
The position is then found by solving the Lorentz Equation (\ref{lorentz}).
In the second approach, the magnetic field is generated for a given volume at the beginning of the
simu\-lation and then it evolves according to Equation (\ref{omega}).
In order to have an acceptable $k$-binning with a dynamical range of $k_{max} / k_{min}  = 10^{10}$,
one would then need to specify the field at an excessively large number of lattice points.
This is not only time-consuming, but also very memory-intensive.
So, like Giacalone \& Jokipii (1994), we adopt the former approach. In
this way, the magnetic field is generated only where needed, and
the overwhelming amount of computer memory required by the second approach is not necessary,.
Since the confinement volume is a parameter of the model,
the second approach would also require adapting the lattice spacing
in order to maintain the same space resolution in physical space.

The Runge-Kutta integrator has previously been vali\-dated
for the cases of uniform and constant electric and magnetic fields,
where the outcome of the simulation can be compared with an analytical solution.

Secondly, as a validation test of the code, we reproduced the result of Giacalone
\& Jokipii (1994) for
the case of a 3D magnetostatic turbulence, by using the same set of para\-meters.
We discretize the turbulence in 50 transverse modes $k$,
where the values of $k$ are chosen to be evenly spaced in logarithmic scale
in the interval of the corresponding length scales from $l_{min} = (1/5)\,v_0/\Omega_0$ to
$l_{max} = 10\,v_0/\Omega_0$, where $v_0$ is the initial velocity of the
particle and $\Omega_0$ is its gyrofrequency in the background magnetic field.
The particles are released in a random initial position
with initial velocity randomly oriented but with fixed value $v_0$.
In Figure~\ref{GJ}, we present the trajectory of a single particle,
the position along the three axes expressed in units of the gyration radius.
In this test, the energy of the particle is constant within a relative error of $10^{-5}$
over a time interval corresponding to $10^3 \Omega_{0} ^{-1}$.

In order to produce the time-dependent turbulent magnetic field considered in this paper,
between two successive shufflings of all five random quantities in $\delta \vect{\Omega}$,
which are performed every $\Delta t \sim 10^8 - 10^9$s,
the particle propagates gyroresonantly with the oscillating turbulence.
We verified that a change in the Runge-Kutta time-binning by over one order of magnitude
does not produce any systematic numerical effects associated with
$\gamma(t)$ and the spectrum. Changing the $k$-binning from $N=1200$ to $N=3000$
in Equation \ref{omega} similarly does not noticeably change the resultant
$\gamma(t)$ and the spectrum. We chose $N=2400$ which results in a reasonably long
computational time. However, a coarser $k$-binning, e.g.,
with $10$ modes/decade, could possibly result in a worse determination of the
macroscopic indicator as instantaneous spatial diffusion coefficients;
the time evolution of the diffusion coefficients is however beyond the scope of the present paper.

\begin{figure}
\begin{center}
\includegraphics[width=8cm]{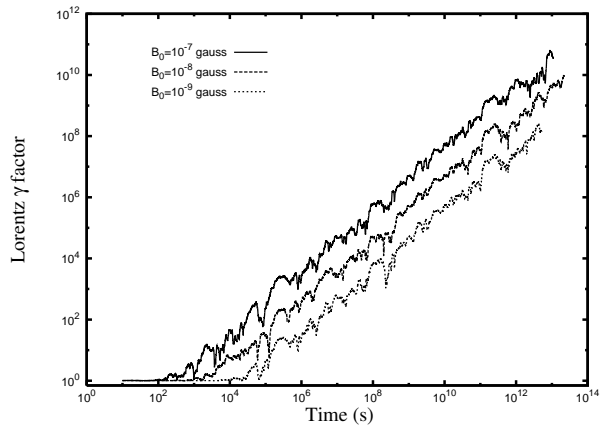}
\caption{Simulated time evolution of the Lorentz factor $\gamma$
for a proton propagating through a time-varying turbulent magnetic field.
The three curves correspond to three different values of $B_0$: $10^{-7}$,
$10^{-8}$, and $10^{-9}$ gauss. The protons are released
at an initial random position inside
the acceleration zone---a sphere of radius ${\mathcal R} = 50$ kpc---with
the same initial speed $v_0$, though pointed in random directions.
The proton is followed until it leaves the acceleration zone and enters
the intergalactic medium. The acce\-leration timescale $\Delta t$ is inversely
proportional to the background field $B_0$. Therefore, as expected, a larger
$B_0$ produces a more efficient acceleration. In this example, a proton
winding its way through a field $B_0 = 10^{-8}$ gauss attains an energy
$E \sim 10^{20}$ eV in approximately $10^6$ years.}
\label{gamma2}
\end{center}
\end{figure}

\section{Results and discussion}
Figure~\ref{part} (to be compared with Figure~\ref{GJ})
shows the trajectory of a single particle released at
random within the acce\-leration zone, with the initial speed $v_0$,
pointed in random direction, with an ambient magnetic field $B_{0} = 10^{-8}$ gauss.
In Figure~\ref{gamma2}, we plot the time evolution of the particle Lorentz factor $\gamma$
for three representative values of the background field $B_0$: $10^{-7}$,
$10^{-8}$, and $10^{-9}$ gauss. We see the particle undergoing various phases of
acceleration and deceleration as it encounters fluctuations in $\vect{B}$.

The acceleration of the particle results from the 0-$th$ component of Equation \ref{geo}, which reads
\begin{equation}
\frac{d\gamma}{dt} = \frac{\delta \vect{\cal E}^i u_i}{\gamma},
\label{gamma0}
\end{equation}
where $\delta \vect{\cal E}^i u_i$ is the scalar product of the electric field and the 3-velocity of the particle.
Therefore the acceleration is given by
$\gamma (t) = \sqrt{{\gamma_0}^2 + 2 \int_{t_0} ^t  \delta \vect{\cal E}^i u_i dt'}$. The integrand function can be strongly time-varying and therefore needs to be computed numerically.

As can be seen in Figure~\ref{gamma2}, the acceleration timescale $\Delta t_{acc}$ is inversely
proportional to $B_0$ so, as expected, more energetic turbulence accelerates the particles
more efficiently, in agreement with what was expected from the non-relativistic
Alfv\'en wave theory. Previous studies (Casse et al. 2002) of particle transport through
a turbulent magnetic field using the prescription in Giacalone \& Jokipii (1994),
compared with a Fast Fourier Transform method,
showed that the time of confinement within the jets of FR II galaxies is
too small for the particles to attain an energy of $10^{20}$ eV. In our case,
the particle acceleration takes place over a much bigger volume
(with dimension ${\mathcal R} = 50$ kpc, compatible with the known size of FR-II radio galaxies)
and the efficiency is enhanced by the strong temporal variation of the turbulence.
We find, in particular, that particles can easily accelerate to UHE on timescales short
compared to the age of the radio-lobe structure through a gyroresonant interaction
with a magnetic turbulence.

In our simulation, the particle acceleration is efficient because it occurs over
a wide range of turbulent fluctuations, such that the wave-particle interaction
is resonant at all times. Such a distribution is expected if the magnetic
energy cascade proceeds (without loss) from the largest spatial scales down
to the region where energy dissipation and transfer to the particles becomes
most efficient.

\begin{figure}
\begin{center}
\includegraphics[width=8cm]{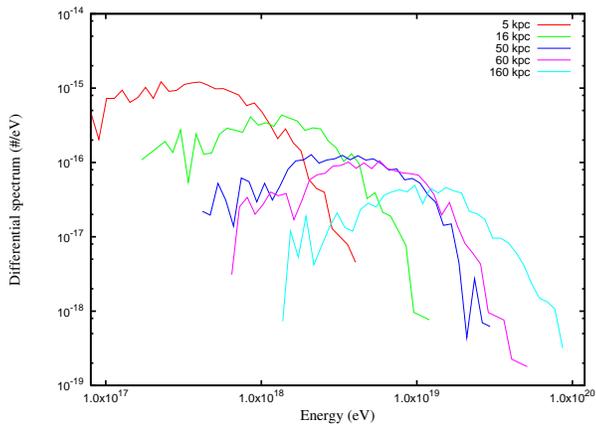}
\caption{Calculated differential spectra for 500 protons with $B_0 = 10^{-8}$ gauss
and for different values of the size of the acceleration region, assumed to be a sphere of radius spanning
the interval ${\mathcal R}=[5-160]$ kpc. The dependence of the energy cutoff on ${\mathcal R}$
is evidenced. This result shows that the cutoff in the observed spectral distribution can be due
to the competition between two distint effects: propagation through the CMB
and intrinsic properties of the accelerator.
Moreover, the slope in the region $E > 4 \times 10^{18}$ eV strongly depends
on $R$. This diagram supports the view that the steeper CR spectrum
below $\log(E/eV) \approx 18.6$
likely represents a population of galactic cosmic rays.}
\label{histo}
\end{center}
\end{figure}

\begin{figure}
\begin{center}
\includegraphics[width=8cm]{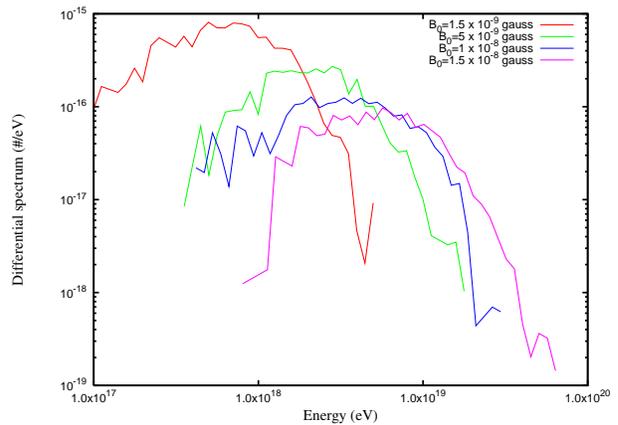}
\caption{Calculated differential spectra for 500 protons with $R = 50$ kpc
and for different values of the turbulent magnetic energy. In this case $B_0$ spans
the interval $B_0=1.5\times[10^{-9}-10^{-8}]$ gauss. See the comments in Figure \ref{histo}.}
\label{histo2}
\end{center}
\end{figure}

In Figures \ref{histo} and \ref{histo2} we show the spectral distributions for distinct values of
the turbulent energy and size of the acceleration region.
In order to produce this result, we followed the
trajectory of 500 protons, launched in the manner described above, with different values of the parameters
$R$ and $B_0$.

We conclude that the observed spectral cutoff (Abbasi et al. 2008, Auger 2008b) can result from
the competition of two distint effects: not only the GZK cutoff,
namely degradation of primary UHECRs
due to the propagation through the CMB, but also, and possibly dominant,
intrinsic properties of the source which constrain the process of acceleration.

\begin{figure}
\begin{center}
\includegraphics[width=8cm]{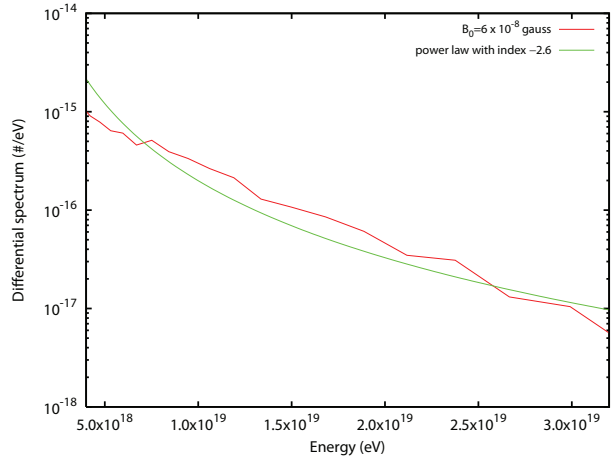}
\caption{Calculated differential spectrum for 1,000 protons in the energy range
$\log(E/eV) = [18.6-19.5]$ for the selected parameters
$B_0 = 10^{-8}$ gauss and ${\mathcal R} = 50$ kpc.
A power-law behaviour with index $-2.6$ in the
differential spectrum of protons injected into the intergalactic medium
in this model is in agreement with a recent statistical analysis of HiRes observations
(Gelmini et al. 2007). This good match supports the view that the steeper CR spectrum
below $\log(E/eV) \approx 18.6$
represents a different population, possibly associated with the Galaxy itself.}
\label{gamma3}
\end{center}
\end{figure}

Figure~\ref{gamma3} depicts the calculated differential spectrum in the energy range
$\log(E/{\rm eV}) = [18.6-19.5]$. From our sampling of the various physical
parameters, we infer that for a radius ${\mathcal R} = 50$ kpc,
$B_0$ should lie in the range $[0.5-5]\times10^{-8}$
gauss in order to produce UHECRs with the observed distribution shown in
this figure.

It is worth emphasizing that this calculation was carried out
without the use of several unknown factors often required in approaches
solving the hybrid Boltzmann equation to obtain the phase-space distribution
function for the particles. In addition, we remark that the acceleration mechanism
we have invoked here is sustained over 10 orders of magnitude in particle energy,
beginning at $\gamma\sim 1$; the UHECRs therefore emerge naturally from the
physical conditions thought to be prevalent within the giant radio lobes of
AGNs, without the introduction of any additional exotic physics (for a
complete review of the bottom-up models,
see Bhattacharjee \& Sigl 2000 and other references cited therein).

\begin{figure}
\begin{center}
\includegraphics[width=8cm]{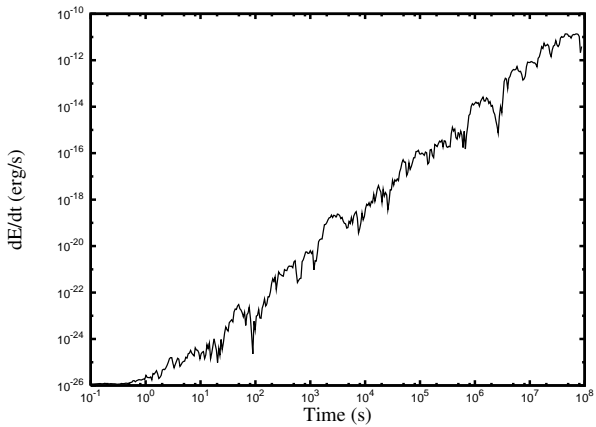}
\caption{Energy loss rate of a single electron in a turbulent time-varying magnetic field
with $B_0 = 10^{-8}$ gauss. In this diagram both the energy losses
due to synchrotron and inverse Compton on the
Radio and CMB photons in the acceleration region are shown. Unlike the case
of protons and heavy ions, the radiation rate for an electron exceeds the
acceleration rate in such a way that, for the given $B_0$,
in a time of order of $10$ years the electron will have lost all of its energy.}
\label{el_loss}
\end{center}
\end{figure}

To provide the possibility of observationally testing the model we have presented here,
we show in Figure \ref{el_loss} the temporal evolution of the
energy loss rate to first order in $\gamma h \nu / (m_e c^2)$ (Blumenthal \& Gould 1970)
for a single electron propagating through the same magnetic turbulence we have used
to accelerate the protons and heavier ions. By estimating the flux of UHE protons 
$\dot N_p$ escaping from one giant radio lobe, under the assumption of neutrality 
in the source, we can estimate the flux of accelerated electrons $\dot N_e$. 
In principle, it is therefore possible to estimate the expected radio luminosity 
from these regions due to this particular acceleration process.

We should point, however, that though a comparison of our results with the
observations supports the viability of this model, our calculations are subject
to several factors we have not fully explored here. For example, the observed
spectrum may be affected by the cosmological evolution in source density (\citet{bgg05}).
However, this omission will not be overly constraining since a likelihood
analysis (\citet{gks07}) of the dependence of the observed distribution
on input parameters has already shown that, in the case of protons,
for $m \sim 0$, where $m$ is the evolution
index in the source density, the HiRes observations are compatible with a
power-law injection spectrum with index $-2.6$. The analysis of the Auger
data seems to confirm this (\cite{auger2008c}).
Thus, although source-density evolution may alter
our results somewhat, our conclusions will probably not be greatly affected.
In a more conservative interpretation, the result presented here provides
the injection spectrum from a single source.

Second, we have not included the GZK effect for particle energies above 50
EeV. This omission becomes progressively more important as the energy approaches
$10^{20}$ eV. These refinements, in addition to a more detailed analysis of the
composition of primary UHECRs, will be reported in a forthcoming paper.
Any discussion concerning the evolution of the instantaneous
spatial diffusion coefficients parallel to and perpendicular to the background
magnetic field, and on the transition to a diffusive regime, will also be
reserved to a future publication.

In our approach, we have also neglected the backreaction of the accelerated particles
on the turbulent field which might increase the ratio $|\delta \vect{\Omega}| / |\vect{\Omega}|$,
and bring about a possible local failure of the assumption of isotropy of the turbulence.

We remark that the mechanism of stochastic acceleration presented here may
be functioning even for a population of particles, protons or heavy nuclei,
pre-accelerated to an initial energy $E \sim 10^{12}- 10^{15}$ eV, e.g.,
by multi-shock fronts propagating at super-Alfv\'enic velocity.
The corresponding gyroresonant wavenumber range in this case
will decrease down to $k_{max}/k_{min} \sim 10^4 - 10^5$.

\section{Conclusion}

We have shown that a region containing a Kolmogorov (turbulent) distribution
of non-relativistic Alfv\'en waves can accelerate particles to ultra-high
energies. The physical
parameters in these regions are compatible with those believed to
be operating in the radio lobes of AGNs. We have discussed the predicted
differential spectrum within the parameter space of the model,
characterized by the size ${\mathcal R}$ of the acceleration region and the
turbulent magnetic energy. Possible tests of this model involve the synchrotron
or IC emission by a population of similarly accelerated electrons.

As the Auger observatory continues to gather more data, improving on the statistics,
our UHECR source identification will continue to get better. Eventually, we should
be able to tell how significant the GZK effect really is, and whether the cutoff
in the CR distribution is indeed due to propagation effects, or whether it is
primarily the result of limitations in the acceleration itself. Given the fact
that energies as high as $\sim 10^{20}$ eV may be reached within typical
radio lobes, it is possible that both of these factors must be considered in
future refinements of this work.

\section*{Acknowledgments}
{\footnotesize\noindent
FF thanks J. Kirk, D. Semikoz, Y. Gallant and A. Markowith for useful and stimulating discussions.
The work of FF was supported by CNES (the French Space Agency) and was carried
out at Service d'Astrophysique, CEA/Saclay and partially at
Laboratory Universe and Theories (LUTh) in
Observatoire de Paris-Meudon and at the  Center for
Particle Astrophysics and Cosmology (APC) in Paris.
This research was partially supported by NSF grant 0402502 at the
University of Arizona. Part of this work was carried out at
Melbourne University and at the Center for Particle Astrophysics
and Cosmology in Paris.}

\label{lastpage}

\end{document}